%%%%%%%%%%%%%%%%%%%%%%%%%%%%%%%%%%%%%%%%%%%%%%%%%%%%%%%%%%%%
%%%%%%%%%%%%%%%%%%%%%%%%%%%%%%%%%%%%%%%%%%%%%%%%%%%%%%%%%%%%
%%%  Instability, Isolation, and the Tridecompositional 
%%%  Uniqueness Theorem.
%%%
%%%  quant-ph/0412013
%%%  December 2004
%%%
%%%   Plain TeX, 20 pages
%%%
%%%   Matthew J. Donald
%%%
%%%   web site:  http://www.poco.phy.cam.ac.uk/~mjd1014
%%%
%%%   e-mail  :  matthew.donald@phy.cam.ac.uk
%%%%%%%%%%%%%%%%%%%%%%%%%%%%%%%%%%%%%%%%%%%%%%%%%%%%%%%%%%%%
%%%%%%%%%%%%%%%%%%%%%%%%%%%%%%%%%%%%%%%%%%%%%%%%%%%%%%%%%%%%

\magnification=1200
\hsize=13cm
\def\newline{\hfil\break}  \def\newpage{\vfil\eject}
\def\proclaim#1#2{\medskip\noindent{\bf #1}\quad \begingroup #2}
\def\endproclaim{\endgroup\medskip}
\def\proof{\noindent{\sl proof}\quad}
\def\<{{<}} \def\>{{>}}
\def\tr{{\rm tr}}
\def\H{{\cal H}} 

\def\dsize{\displaystyle}
\abovedisplayskip=3pt plus 1pt minus 1pt
\belowdisplayskip=3pt plus 1pt minus 1pt
\def\hcrh{\hfill \cr \hfill} \def\crh{\cr \hfill} \def\hcr{\hfill \cr}
\def\blacksquare{\vrule height 4pt width 3pt depth2pt}

\font\brm=cmbx12  \def\tilbf{\lower 1.1 ex\hbox{\brm \char'176}}

\font\trm=cmr12  
\def\tilrm{\lower 1.1 ex\hbox{\trm \char'176}}

\topskip10pt plus40pt
\clubpenalty=30

\headline={\hfil}
\footline={\hss\tenrm\folio\hss}

{\bf
\centerline{\ \  Instability, Isolation, and the Tridecompositional Uniqueness
Theorem.
\footnote{*}{\tenrm quant-ph/0412013, December 2004.}}
\medskip

\centerline{Matthew J. Donald}
\medskip

\centerline{The Cavendish Laboratory,  Madingley Road,}  

\centerline{Cambridge CB3 0HE,  Great Britain.}
\smallskip

\centerline{ e-mail: \quad matthew.donald@phy.cam.ac.uk}
\smallskip

{\bf \hfill web site:\quad  
{\catcode`\~=12 \catcode`\q=9
http://www.poco.phy.cam.ac.uk/q~mjd1014
}\hfill }}
\bigskip

The tridecompositional uniqueness theorem of Elby and Bub (1994) shows that a
wavefunction in a triple tensor product Hilbert space has at most one
decomposition into a sum of product wavefunctions with each set of component
wavefunctions linearly independent.  I demonstrate that, in many
circumstances, the unique component wavefunctions and the coefficients in
the expansion are both hopelessly unstable, both under small changes in global
wavefunction and under small changes in global tensor product structure.   In
my opinion, this means that the theorem cannot underlie law-like solutions to
the problems of the interpretation of quantum theory.  I also provide examples
of circumstances in which there are open sets of wavefunctions
containing no states with various decompositions.

\proclaim{1. Introduction.}
\endproclaim

The central problem of the interpretation of quantum theory is to explain
and characterize the existence, or apparent existence, of state ``collapse''. 

State collapse is the process by which, for example, an electron, despite
apparently going through a double slit as an extended wave, always appears to
make a well-localized impact on a screen at only one of many possible places --
the extended wave ``collapses'' to a localized state.  Different
interpretations of quantum theory suggest different ways of
understanding such collapses.  One plausible goal would be to propose the
existence of laws of nature defining the circumstances in which collapse
occurs or appears to occur.  I have attempted this myself based on a
characterization of observers and their information (Donald 1999).  To
propose such laws, whether or not they involve observers, is to propose a
realist interpretation of quantum theory, in the sense that the laws are
supposed to be truths about reality which are independent of our abilities
to verify them.

There are many proposals simpler than my own.  Perhaps the simplest is
implicit in a not uncommon understanding of decoherence theory.  This
assumes that, as a consequence of decoherence, a global quantum state just
falls naturally into a family of quasi-classical pieces, each of which
describes the observation of an individual collapse outcome.  The problem
with this is that decoherence theory does not by itself solve the preferred
basis problem.  It merely provides a framework within which a quasi-classical
solution to the preferred basis problem is not ruled out.  Typical quantum
states for macroscopic objects can be split into quasi-classical pieces in
many ways which, at least at a level of fine structure, are mutually
incompatible.  But a realist explanation of collapse along these lines
requires specification of one fundamental splitting.  Which individual piece
we see may be a matter of probability, but to define probabilities we need to
be given a splitting into a definite set of possible pieces.  Although there
are models which provide asymptotically unambiguous decompositions if we wait
long enough (Hepp 1972) or look on a broad enough scale (Ollivier, Poulin, and
Zurek 2003, 2004), such models are not sufficient.  In particular, if we try
to analyse the detailed real-time functioning of an individual human brain,
there is considerable ambiguity as to the precise scales on which information
is being experienced (Donald 2002). 

Another approach supposes that a natural splitting of a global quantum state
might be a consequence of some property of the mathematics of quantum states. 
The paradigm is the Schmidt, or biorthogonal, decomposition.  If the global
Hilbert space $\H$ splits naturally into a tensor product $\H = \H_1 \otimes
\H_2$ and if the global state is a pure state, then its wavefunction $\Psi$ has
a decomposition of the form $\Psi = \sum_k \sqrt{p_k} \psi^1_k \psi^2_k$, where
$p_k \geq 0$ for all $k$, $\sum_k p_k = 1$, and $(\psi^1_k)$ and $(\psi^2_k)$
are orthonormal bases for $\H_1$ and $\H_2$, respectively.  Generically, the
Schmidt decomposition is unique.  It fails to be unique at points of
degeneracy, which are precisely those points where the $p_k$ are not all
distinct.

The idea of using this decomposition to explain collapse or the
appearance of collapse is the idea that $\Psi$ will correspond to an
``extended wave'' of some form, satisfying a global Schr\"odinger equation
of some form, and the component states $\psi^1_k \psi^2_k$ will correspond to
observed collapse states.  The great advantage of this idea is that the
Schmidt decomposition is well-defined.  If a global quantum state is given,
then, generically, its Schmidt decomposition is determined.  Good
mathematical definitions are the required underpinnings for realist laws of
nature.  A law stating that $\Psi$ will collapse or appear to collapse to
component $\psi^1_k \psi^2_k$ with probability $p_k$ is at the heart of the
modal interpretation (or some versions of it).  Bub (1997) provides a brief
review and Vermaas (2000) a detailed examination.

A well-defined law faces questions.  Among the questions which arise for
the Schmidt decomposition are:

\item{1.1)}  How are the subsystems $\H_1$ and $\H_2$ to be identified?

\item{1.2)}  Is it appropriate to assume that the state on $\H_1
\otimes \H_2$ is pure?

\item {1.3)} Given plausible global wavefunctions $\Psi$ on $\H_1 \otimes
\H_2$, are the component wavefunctions $\psi^1_k \psi^2_k$ appropriately
quasi-classical?

\item {1.4)} What about degeneracy points? 

In my opinion, the modal interpretation fails because none of these questions
can be given entirely satisfactory answers.  The literature is extensive and I
shall not review it here.  For the moment, I merely note that, for macroscopic
systems, the answer to question 1.3 seems to be that the component
wavefunctions can be quite arbitrary and can be unstable under small changes
in $\Psi$ and under small changes in the tensor product structure
(Bacciagaluppi, Donald,  and Vermaas 1995, Donald 1998).

The tridecompositional uniqueness theorem, stated in various versions in
section 2, was developed as a tool to solve problems with the Schmidt
decomposition.  It has been invoked to explain the states adopted, or
apparently adopted, by quantum systems in contact with both a
measuring device and an environment (Bub 1997, Schlosshauer 2003).  The
purpose of this paper is to show that it has its own problems.  Instability
problems will be exemplified in section 3, culminating in theorems 3.4 and
3.6.  In section 4, it will be shown that there are open sets, and other
significant sets, which contain no wavefunctions with various decompositions. 
In section 5, in a counterpoint to the central thrust of the paper, detailed
technical estimates will be used to show that triorthogonal decompositions --
those satisfying theorem 2.3 -- are in fact stable.  The results can be
summarized by saying that, in large spaces, general tridecompositions tend to
exist but are unstable, while triorthogonal decompositions are stable but
unlikely. 

For brevity, it will be assumed throughout that all named Hilbert spaces are
non-trivial, and in particular that they have dimension at least two.
\medskip

\proclaim{2. The Tridecompositional Uniqueness Theorem.}
\endproclaim

\proclaim{Theorem 2.1}{\sl}  Let $\H = \H_1 \otimes \H_2 \otimes \H_3$
be a triple tensor product of Hilbert spaces and let $\Psi \in \H$ be a
wavefunction.

Then $\Psi$ has at most one decomposition of the form
$\Psi = \sum_{k = 1}^K a_k \psi^1_k \psi^2_k \psi^3_k$ \newline where $K$ is
finite, \
$|a_k| > 0$ for $k = 1, \dots, K$, \
$\{\psi^1_k: k = 1, \dots, K\} \subset \H_1$ and \newline $\{\psi^2_k:
k = 1,
\dots, K\}  \subset \H_2$ are linearly independent sets of wavefunctions, and
\newline $\{\psi^3_k: k = 1, \dots, K\} \subset \H_3$ is a set of wavefunctions
in
$\H_3$ such that no pair is collinear.
\endproclaim

Note that  $\psi^1_k \psi^2_k \psi^3_k$ abbreviates $\psi^1_k \otimes
\psi^2_k \otimes \psi^3_k$, that a wavefunction $\psi$ is a normalized vector
($||\psi|| = 1$), that a finite set $\{\psi_k: k = 1, \dots, K\}$ of
wavefunctions is linearly independent iff 
$\sum_{k = 1}^K c_k \psi_k = 0$ implies $c_k = 0$ for all $k$, and that a pair
$\{\psi_k, \psi_l\}$ of wavefunctions is collinear, iff it is not linearly
independent, iff $\psi_k$ and $\psi_l$ differ at most by a phase factor.  The
uniqueness of the decomposition in this theorem, of course, allows for
re-orderings of the terms and changes in phase factors.

Theorem 2.1 is proved in Elby and Bub (1994), with the argument improved and
completed in Clifton (1994) and Bub (1997).  Kirkpatrick
(2001) extends the result by noting that it is not necessary to specify which
particular pair of spaces have linearly independent wavefunctions. 
Cassam-Chena{\"\i} and Patras (2004) set the theorem in a powerful
algebraic framework allowing them to consider spaces of indistinguishable
particles.  In section 3, I shall show that, in many circumstances, although
the decomposition in theorem 2.1 is unique at individual wavefunctions, it
can vary wildly as we move from wavefunction to wavefunction.  The
fundamental source of this instability is that linear independence is a very
weak property.  If there are enough spare dimensions available, then 
arbitrarily small modifications can turn a finite set of wavefunctions into a
linear independent set.

It will be useful to be able to refer to the statements of the following
consequences of theorem 2.1.  Elby and Bub have named theorem 2.3
the ``triorthogonal uniqueness theorem''.  We shall refer to a decomposition
satisfying theorem 2.3 as a ``triorthogonal decomposition'', and to a
wavefunction which has a triorthogonal decomposition as a ``triorthogonal
wavefunction''.

\proclaim{Theorem 2.2}{\sl}  Let $\H = \H_1 \otimes \H_2 \otimes \H_3$
be a triple tensor product of Hilbert spaces and let $\Psi \in \H$ be a
wavefunction.

Then $\Psi$ has at most one decomposition of the form
$\Psi = \sum_{k = 1}^K a_k \psi^1_k \psi^2_k \psi^3_k$ where $K$ is finite, $|a_k| >
0$ for $k = 1, \dots, K$, and, for $i = 1, 2, 3$,
$\{\psi^i_k: k = 1, \dots, K\}$ is a linearly independent set of
wavefunctions in $\H_i$.
\endproclaim

\proclaim{Theorem 2.3}{\sl}  Let $\H = \H_1 \otimes \H_2 \otimes \H_3$
be a triple tensor product of Hilbert spaces and let $\Psi \in \H$ be a
wavefunction.

Then $\Psi$ has at most one decomposition of the form
$\Psi = \sum_{k = 1}^K a_k \psi^1_k \psi^2_k \psi^3_k$ where $|a_k| >
0$ for $k = 1, \dots, K$, and, for $i = 1, 2, 3$,
$\{\psi^i_k: k = 1, \dots, K\}$ is a orthonormal set of wavefunctions in
$\H_i$.
\endproclaim

In the statements given here, theorem 2.1 and 2.2 refer only to finite
decompositions.  There are two reasons for this.  Firstly the conventional
definition of linear independence states that an infinite set of wavefunctions
is linearly independent if and only if every finite subset is linearly
independent.  For infinite sets $\{\psi_k\}$, this is weaker than the
statement that $\sum_k c_k \psi_k = 0$ implies $c_k = 0$.  Secondly, the
proofs of the theorems involve expanding one linearly independent set in
terms of another.  For infinite sets, the coefficients of these expansions
can, in general, become unbounded.

These problems are not relevant in the case of theorem 2.3.  In that case, the
decomposition also amounts to a Schmidt decomposition of $\Psi$ in the tensor
product of $\H_1$ with $\H_2 \otimes \H_3$.  This uniquely identifies the
$|a_k|$ and, for each $\delta > 0$, the finite dimensional space spanned by the
$\psi^1_k \psi^2_k \psi^3_k$ with $|a_k| > \delta$.  Theorem 2.3 then follows
from the uniqueness of decomposition on these finite spaces, which is a
consequence of theorem 2.1.
\medskip

\proclaim{3. Instability.}
\endproclaim

\proclaim{Example 3.1}
 Let $\H = \H_1 \otimes \H_2 \otimes \H_3$ where, for $i = 1, 2, 3$,
$(\psi_n^i)_{n =1}^{N_i}$ is  an orthonormal basis for $\H_i$.  Suppose that
$\Psi = {\textstyle{1\over \sqrt{2}}}(\psi^1_1  \psi^2_1 
\psi^3_2 - \psi^1_1  \psi^2_2  \psi^3_1)$.

$\Psi$ is a tensor product of $\psi^1_1$ with a singlet wavefunction $
{\textstyle{1\over \sqrt{2}}}(\psi^2_1  \psi^3_2 - \psi^2_2  \psi^3_1)
\in \H_2 \otimes \H_3$.

One of the original aims of Elby and Bub (1994) in developing the
tridecompositional uniqueness theorem was to avoid problems which are
raised by the non-unique Schmidt decomposition of singlets.  This
same non-uniqueness, however, can be used to provide an example of
instability in the wavefunctions.

Write $\varphi^2_1 = {1\over \sqrt{2}}(\psi^2_1 - \psi^2_2)$ and
$\varphi^2_2 = {1\over \sqrt{2}}(\psi^2_1 + \psi^2_2)$ and write
$\varphi^3_1 = {1\over \sqrt{2}}(\psi^3_1 + \psi^3_2)$ and
$\varphi^3_2 = {1\over \sqrt{2}}(\psi^3_1 - \psi^3_2)$.

Then
$$\Psi = {\textstyle{1\over \sqrt{2}}}(\psi^1_1  \varphi^2_1  \varphi^3_1 -
\psi^1_1  \varphi^2_2  \varphi^3_2).$$

Let $\varphi^1_1 = \xi^1_1 = \psi^1_1$ and $\varphi^1_2(\theta) = \xi^1_2(\theta) = - \cos \theta
\psi^1_1 - \sin \theta \psi_2^1$.  Note that $\{\psi^1_1, - \cos \theta
\psi^1_1 - \sin \theta \psi_2^1\}$ is linearly independent as long as $\sin \theta \ne
0$.

Write $\xi^2_1 = \psi^2_1$, $\xi^2_2 = \psi^2_2$, $\xi^3_1 =
\psi^3_2$, and $\xi^3_2 = \psi^3_1$.

For $0 < \theta \leq {\pi \over 2}$,
$$\Phi(\theta) = {\textstyle{1\over \sqrt{2}}}(\varphi^1_1  \varphi^2_1  \varphi^3_1
+ \varphi_2^1(\theta)  \varphi^2_2  \varphi^3_2)$$
and
$$\Psi(\theta) = {\textstyle{1\over \sqrt{2}}}(\xi^1_1  \xi^2_1  \xi^3_1
+ \xi_2^1(\theta)  \xi^2_2  \xi^3_2)$$
both satisfy the conditions of the theorem 2.2, and so these expansions are
unique and $\Phi(\theta) \ne \Psi(\theta)$.  However
$$\lim_{\theta \rightarrow 0} \Phi(\theta) = \lim_{\theta \rightarrow 0}
\Psi(\theta) = \Psi$$ and so the components of the terms in the expansion are
not stable.
\hfill $\blacksquare$
\endproclaim

The use of a singlet is not critical in this example.  Indeed, any
wavefunction $\psi^{23}$ on $\H_2 \otimes \H_3$ which is not a product can
be decomposed into a sum of products with linearly-independent
components in many different ways.
Such a wavefunction will have a Schmidt decomposition of the form
$\psi^{23} = \sum_{k = 1}^K \sqrt{p_k} \psi^2_k \psi^3_k$ with $p_1 \geq p_2 > 0$. 
This can be written in different ways by using the identity
$$\displaylines{
 \sqrt{p_1} \psi^2_1 \psi^3_1 + \sqrt{p_2} \psi^2_2 \psi^3_2 
= (\cos\alpha \psi^2_1 + \sin \alpha \psi^2_2)(\sqrt{p_1} \cos\alpha \psi^3_1 +
\sqrt{p_2} \sin\alpha \psi^3_2)
\hcrh + (\sin\alpha \psi^2_1 - \cos\alpha \psi^2_2)(\sqrt{p_1} \sin\alpha \psi^3_1 -
\sqrt{p_2} \cos\alpha \psi^3_2).   }$$

Instability in tridecompositions close to $\psi^1_1 \psi^{23}$ follows as in
example 3.1.

\proclaim{Example 3.2}
Kirkpatrick (2001) has used theorem 2.1 to prove that a state $\rho$ on a
tensor product space $\H_1 \otimes \H_2$ has at most one convex
decomposition of the form $\rho = \sum_{k = 1}^K w_k |\psi^1_k \psi^2_k\>\<
\psi^1_k \psi^2_k|$ where $K$ is finite, $w_k > 0$ for $k = 1, \dots, K$,
and one of the sets $\{\psi^1_k: k = 1, \dots, K\} \subset \H_1$ and
$\{\psi^2_k: k = 1, \dots, K\}  \subset \H_2$ is linearly independent while, in
the other, no pair of wavefunctions is collinear.  Example 3.1 can be extended
to show that the components of Kirkpatrick's decomposition are also unstable.

As $\{\varphi^3_1, \varphi^3_2\}$ and $\{\xi^3_1, \xi^3_2\}$ are orthonormal,
the partial traces of $|\Phi(\theta)\>\<\Phi(\theta)|$ and
$|\Psi(\theta)\>\<\Psi(\theta)|$ on $\H_1 \otimes \H_2$ are given by
$$(|\Phi(\theta)\>\<\Phi(\theta)|)_{12} =  \textstyle{1\over 2}|\varphi^1_1 
\varphi^2_1\>\<\varphi^1_1  \varphi^2_1| +
 {1\over 2}|\varphi_2^1(\theta)  \varphi^2_2\>\<\varphi_2^1(\theta) \varphi^2_2|$$
and
$$(|\Psi(\theta)\>\<\Psi(\theta)|)_{12} =  \textstyle{1\over 2}|\xi^1_1 
\xi^2_1\>\<\xi^1_1  \xi^2_1| +
 {1\over 2}|\xi_2^1(\theta)  \xi^2_2\>\<\xi_2^1(\theta) \xi^2_2|.$$

For $0 < \theta \leq {\pi \over 2}$, these decompositions satisfy the conditions of
Kirkpatrick's theorem, so that the components are unique.  However,
although, as $\theta \rightarrow 0$,
$$||(|\Phi(\theta)\>\<\Phi(\theta)|)_{12} - (|\Psi(\theta)\>\<\Psi(\theta)|)_{12}|| \rightarrow 0,$$
the components do not converge.  Indeed, for all $\theta$,
$|\<\varphi^1_i  \varphi^2_i|\xi^1_j  \xi^2_j\>| \leq {1 \over \sqrt{2}}$ for $i, j \in \{1,
2\}$.
\hfill $\blacksquare$
\endproclaim

\proclaim{Example 3.3}
Once again, let $\H = \H_1 \otimes \H_2 \otimes \H_3$ where, for $i = 1, 2,
3$, $(\psi_n^i)_{n =1}^{N_i}$ is  an orthonormal
basis for $\H_i$.  This time however take $\Psi = \psi^1_1  \psi^2_1 \psi^3_1$.

Clearly, for any constant $X$, $\Psi = (1 - X) \psi^1_1  \psi^2_1 \psi^3_1 + X
\psi^1_1 \psi^2_1 \psi^3_1$.

Write $\varphi^1_1 = \psi^1_1$, $\varphi^2_1 = \psi^2_1$, $\varphi^3_1 = \psi^3_1$ and
$\varphi^1_2(\theta) =  \cos \theta \psi^1_1 + \sin \theta \psi^1_2$,
$\varphi^2_2(\theta) =  \cos \theta \psi^2_1 + \sin \theta \psi^2_2$, and
$\varphi^3_2(\theta) =  \cos \theta \psi^3_1 + \sin \theta \psi^3_2$.

For $0 < \theta \leq {\pi \over 2}$, let $\Phi(\theta) = (1 -
1/\sqrt{\theta}) \varphi^1_1  \varphi^2_1 \varphi^3_1 + (1/\sqrt{\theta})
\varphi_2^1(\theta) \varphi^2_2(\theta) \varphi^3_2(\theta)$ and
$\Psi(\theta) = \Phi(\theta)/||\Phi(\theta)||$.
$||\Phi(\theta)||$ is bounded and tends to 1 as $\theta \rightarrow 0$.

The expansion for $\Psi(\theta)$
satisfies the conditions of theorem 2.2, but, although $\Psi(\theta)
\rightarrow \Psi$, the coefficients of the expansion diverge as $\theta
\rightarrow 0$. \hfill $\blacksquare$
\endproclaim

\proclaim{Theorem 3.4}{\sl} In a triple product of infinite-dimensional
spaces, arbitrarily close to every wavefunction are pairs of wavefunctions
satisfying the conditions of the tridecompositional theorem with
arbitrarily different component wavefunctions.

Specifically, let $\H = \H_1 \otimes \H_2 \otimes \H_3$ be a triple product
of infinite-dimensional spaces and let
$\Psi \in \H$ be a wavefunction.  Choose $\varepsilon > 0$ and,  for $i =
1, 2, 3$, choose an orthonormal basis $(u_n^i)_{n\geq1}$ for $\H_i$. 

Then, there exist wavefunctions $\Phi_1$ and
$\Phi_2$ in $\H$ which satisfy the conditions of the  theorem 2.2 with unique
finite expansions 
$$\Phi_1 = \sum_{k=1}^K a_k \psi^1_k \psi^2_k \psi^3_k \hbox{\quad
and \quad}
\Phi_2 = \sum_{m=1}^M b_m \varphi^1_m \varphi^2_m \varphi^3_m$$
such that
$||\Psi - \Phi_1|| < \varepsilon$, $||\Psi - \Phi_2|| < \varepsilon$ and, for $k = 1, \dots, K$, $m =
1, \dots, M$, and $i = 1, 2, 3$, there exists $n(k)$ such that
$|\<\psi^i_k|u^i_{n(k)}\>| > 1 - \varepsilon$ while
$|\<\psi^i_k|\varphi^i_m\>| < \varepsilon$.

Thus each component wavefunction of the expansion of $\Phi_1$ is close to some
element of a freely chosen basis while being far from every component
wavefunction of the expansion of $\Phi_2$.
\endproclaim

\proof

Let $\H$, $\Psi$, $\varepsilon$, and $(u_n^i)_{n\geq1}$ satisfy the hypothesis
of the theorem.

Let $P^N = \sum_{n_1=1}^N \sum_{n_2=1}^N \sum_{n_3=1}^N |u^1_{n_1} u^2_{n_2}
u^3_{n_3}\>\<u^1_{n_1} u^2_{n_2} u^3_{n_3}|$.

$P^N \buildrel s \over \rightarrow 1$.  Choose $N_0$ such that $N \geq N_0$ implies
$||P^N \Psi|| > 0$ and $||\Psi - \Phi^N|| < \varepsilon/2$ where $\Phi^N =  P^N \Psi/||P^N
\Psi||$.

\proclaim{Lemma}  Let $(u_n)_{n=1}^N$ be an orthonormal basis for an
$N$-dimensional Hilbert space.  Then there exists another orthonormal basis
$(v_n)_{n=1}^N$ such that for all pairs $k$ and $m$, 
$|\<u_k|v_m\>| = {1 \over \sqrt{N}}$
\endproclaim

\proof
Define $v_m = {1 \over \sqrt{N}} \sum_{k=1}^N e^{2\pi i k m/ N} u_k$
for $m = 1, \dots, N$.

Then $||v_m||^2 = 1$, and, for $m \ne n$, $$\<v_n|v_m\> =
{{1\over N}} \sum_{k=1}^N e^{2\pi i k (m-n)/ N} = 0.$$

\ \hfill $\blacksquare$

Choose $N \geq N_0$ so that ${1 \over \sqrt{N}} < \varepsilon/2$.

For $i = 1, 2, 3$, define $(v^i_n)_{n=1}^N$ in terms of
$(u^i_n)_{n=1}^N$ as in the lemma.

Suppose that 
$$\Phi^N = \sum_{n_1=1}^N \sum_{n_2=1}^N \sum_{n_3=1}^N a_{n_1, n_2,
n_3}u^1_{n_1} u^2_{n_2} u^3_{n_3} 
=  \sum_{n_1=1}^N \sum_{n_2=1}^N \sum_{n_3=1}^N b_{n_1, n_2, n_3} v^1_{n_1}
v^2_{n_2} v^3_{n_3}  \eqno{(3.5)}$$  
are eigenvector expansions.

Omitting terms with zero coefficients and choosing some ordering for the
remaining terms, these expansions can be re-written in the form
$$\Phi^N = \sum_{k=1}^K a_k  \psi^1_k(0)  \psi^2_k(0)  \psi^3_k(0)
=  \sum_{m=1}^M b_m  \varphi^1_m(0)  \varphi^2_m(0)  \varphi^3_m(0)$$  
where each $\psi^i_k(0) \in \{u^i_n : n = 1, \dots, N\}$ and each 
$\varphi^i_m(0) \in \{v^i_n : n = 1, \dots, N\}$.

These expansions are different, and so they cannot both satisfy the
conditions of the tridecompositional theorem.  Indeed, in general, neither
expansion will, as we will have $K = M = N^3$ and the component
wavefunctions will be repeated.

However, as in example 3.1, an arbitrarily slight perturbation in each
component is sufficient to produce linear independence. 

Let $ \psi^i_k(\theta) = \cos \theta \psi^i_k(0) + \sin \theta\, u^i_{N + 1 +
k}$ for $i = 1, 2, 3$ and $k = 1, \dots, K$ and let
$\varphi^i_m(\theta) = \cos \theta \varphi^i_m(0) +
\sin \theta\, u^i_{N + 1 + m}$ for $i = 1, 2, 3$ and $m = 1, \dots,
M$.

Then, for $0 < \theta \leq {\pi \over 2}$,
$\Phi^N_1(\theta) = \sum_{k=1}^K a_k \psi^1_k(\theta) \psi^2_k(\theta)
\psi^3_k(\theta)$ and \newline
$\Phi^N_2(\theta) = \sum_{m=1}^M b_m \varphi^1_m(\theta) \varphi^2_m(\theta)
\varphi^3_m(\theta)$ do both satisfy the conditions of theorem 2.2.

Taking $\theta$ sufficiently small gives the theorem.
\hfill $\blacksquare$
\medskip

The argument following (3.5) in this proof can be applied to any finite
expansion of $\Phi^N$ in $P^N \H$.  Thus, combining the methods of
example 3.3 and theorem 3.4 shows that, in infinite dimensional spaces,
both the components and the coefficients defined by the tridecompositional
uniqueness theorem are utterly unstable everywhere.  This means that the
theorem, although it is remarkable and powerful as a mathematical tool, is
useless for explanations of the quasi-classical nature of collapse outcomes; in
particular as real physical systems have arbitrarily many degrees of freedom,
including those involving virtual photons, which can be touched upon.
\medskip

Theorem 3.4 is concerned with variations in the global wavefunction.  It is
also important, for the question of the physical relevance of the
tridecompositional uniqueness theorem, to recognize that the assumption
that a physical Hilbert space has a fundamental tensor product structure may
well be incorrect.  This recognition can be supported by consideration of the
derived and phenomenological nature of localized particles according to
relativistic quantum field theory, but, even at a less sophisticated level, it
is hard to justify the idea that there are the sort of natural boundaries
which would allow the universe to be divided without ambiguity into a system,
a measuring apparatus, and an environment.

If the tensor product structure is not a fundamental aspect of reality, then
the ultimate laws of nature cannot depend on it.  This means that any
approximate version of those laws which does depend on a
phenomenological assignment of a tensor product structure should be
stable under small variations of that assignment.  The next theorem will
rework theorem 3.4 to show that the decomposition provided by the
tridecompositional uniqueness theorem is not stable under such variations.

Let $\H$ be an infinite dimensional Hilbert space.  One inelegant but
adequate way of defining the structure of a triple product of
infinite-dimensional spaces on $\H$ is simply to label any given
orthonormal basis in the form $(\Psi_{n_1, n_2, n_3})_{n_1 \geq 1, n_2 \geq 1,
n_3 \geq 1}$.  Then $\H \cong \H_1 \otimes \H_2 \otimes \H_3$ where $\H_1$,
for example, corresponds to the space with operators generated by
$\sum_{n_2 \geq 1, n_3 \geq 1} |\Psi_{m_1, n_2, n_3}\>\<\Psi_{n_1, n_2, n_3}|$.

If $U$ is a unitary map on $\H$, then $(U\Psi_{n_1, n_2, n_3})_{n_1 \geq 1, n_2
\geq 1, n_3 \geq 1}$ is an alternative labelled basis and therefore defines an
alternative triple product structure.  A sufficient, but not a necessary
condition, for this second product structure to be close to the first is that
$U$ be appropriately close to $1$ (the identity map).  A strong measure of
proximity -- the trace class norm ($||\ ||_1$) -- will be invoked in the
next theorem. 

We shall say that a unitary map $U$ on $\H$ moves one tensor product
structure $\H = \H_1 \otimes \H_2 \otimes \H_3$ into another $\H = \H'_1
\otimes \H'_2 \otimes \H'_3$ if, for $i = 1, 2, 3$, there are orthonormal bases
$(u^i_n)_{n\geq1}$ (respectively $(\hat u^i_n)_{n\geq1}$) for $\H_i$
(resp.~$\H'_i$) such that, for all triples $(n_1, n_2, n_3)$,
$Uu^1_{n_1} u^2_{n_2} u^3_{n_3}
= \hat u^1_{n_1}\hat u^2_{n_2}\hat u^3_{n_3}$.  We shall
say that the structures differ in trace norm by the infimum of $||U - 1||_1$
over all such $U$.

Using such a unitary map, aspects of the different structures can be
compared by asking how they would differ if bases moved by $U$ were
identified.  For example, if $\psi^1 \in \H_1$ and $\hat \varphi^1 \in \H'_1$,
then
$\<\psi^1|\hat \varphi^1\>$ is undefined because $\H_1$ and $\H'_1$ are
different spaces but $\<\psi^1 u^2_{n_2} u^3_{n_3}|
U^*|\hat \varphi^1 \hat u^2_{n_2} \hat u^3_{n_3}\>$, which is
independent of $n_2$ and of $n_3$ and of the choice of bases moved by $U$, is
an appropriate substitute.  We shall denote this expression by
$\<\psi^1|\hat \varphi^1\>_U$ and will use a similar notation for analogous
expressions.  In fact, the identification of bases defines isomorphisms between
$\H_i$ and $\H_i'$ for $i = 1, 2, 3$.

\proclaim{Theorem 3.6}{\sl}  Let $\H = \H_1 \otimes \H_2 \otimes \H_3$
be a triple product of infinite-dimensional spaces and let
$\Psi \in \H$ be a wavefunction.  Choose $\varepsilon > 0$.  Then:
\smallskip

\noindent (3.7) \  There exists a wavefunction $\Phi_1 \in \H$ with $||\Psi -
\Phi_1|| < \varepsilon$ which satisfies the conditions of theorem 2.2 with a unique
finite expansion 
$\Phi_1 = \sum_{k=1}^K a_k \psi^1_k \psi^2_k \psi^3_k$.

 There also exists another representation $\H = \H'_1 \otimes \H'_2
\otimes \H'_3$ of $\H$ as a triple product of infinite-dimensional spaces,
such that:

\noindent (3.8) \  The two triple product structures differ in trace norm
by less than $4\varepsilon$.

\noindent (3.9) \  In the second structure, $\Phi_1$ also satisfies the
conditions of theorem 2.2 with a unique finite expansion 
$\Phi_1 = \sum_{m=1}^M b_m \hat \varphi^1_m \hat \varphi^2_m \hat \varphi^3_m$.

\noindent (3.10) \  For $k = 1, \dots, K$, $m = 1, \dots, M$, and $i = 1, 2, 3$,
$|\<\psi^i_k|\hat \varphi^i_m\>_U| < \varepsilon$.
\endproclaim

\proof Adopt the notation and definitions of theorem 3.4.  (3.7) follows
immediately. 

As they have different unique expansions, clearly $\Phi_1 \ne \Phi_2$.

$\Phi_1$ and $\Phi_2$ can be written in the forms $\Phi_1 = \alpha \Phi_2 + \beta
\Phi_2^\perp$ and  $\Phi_2 = \bar \alpha \Phi_1 - \beta \Phi_1^\perp$ where 
$\<\Phi_2^\perp|\Phi_2\> = 0$, $\<\Phi_1^\perp|\Phi_1\> = 0$, and $|\alpha|^2
+ |\beta|^2 = 1$.  The choice of phases here implies that
$\<\Phi_1^\perp|\Phi_2^\perp\> = \<\Phi_2|\Phi_1\> = \alpha$ and so
$$||\Phi_1 - \Phi_2||^2 = ||\Phi_1^\perp - \Phi_2^\perp||^2 = 2 - \alpha -
\bar \alpha.$$

Then a unitary map $U$ on $\H$ mapping $\Phi_2$ to $\Phi_1$ can be defined
by 
$$U = |\Phi_1\>\<\Phi_2| + |\Phi_1^\perp\>\<\Phi_2^\perp| + 1 -
|\Phi_1\>\<\Phi_1| - |\Phi_1^\perp\>\<\Phi_1^\perp|.$$

This is the unitary map which the identity except on the space spanned
by $\Phi_1$ and $\Phi_2$ on which it sends $\Phi_2$ to $\Phi_1$ and
$\Phi_2^\perp$ to $\Phi_1^\perp$.

On that space $U -1$ has matrix representation
$U - 1 = \pmatrix{ \alpha - 1 & -\bar \beta \cr \beta & \bar \alpha - 1  \cr}$.

It is clear that $U \rightarrow 1$ as $\Phi_1 \rightarrow \Phi_2$.  
Matrix multiplication shows that 
\newline $(U - 1)^* (U - 1) = (2 - \alpha - \bar \alpha) \pmatrix{ 1
& 0 \cr 0 & 1  \cr}$ from which it follows that
$$||U - 1||_1 = 2\sqrt{2 - \alpha -\bar \alpha} = 2||\Phi_1 - \Phi_2|| < 4\varepsilon.$$
This is the bound in (3.8).

Move the triple product structure $\H = \H_1 \otimes \H_2 \otimes \H_3$
with $U$ to $\H = \H'_1 \otimes \H'_2 \otimes \H'_3$.
For $i = 1, 2, 3$, define bases $(\hat u^i_n)_{n\geq1}$ for $\H_i'$ by
$U |u^1_{n_1} u^2_{n_2} u^3_{n_3}\>
=  |\hat u^1_{n_1}  \hat u^2_{n_2} \hat u^3_{n_3}\>$,
and if $\varphi^i_m = \sum_{n} f^i_{mn} u^i_n$ then define
$\hat \varphi^i_m = \sum_{n} f^i_{mn} \hat u^i_n$.

Suppose that $\Phi_2 =  \sum_{n_1, n_2, n_3 \ge 1} c_{n_1, n_2, n_3}  
u^1_{n_1} u^2_{n_2} u^3_{n_3}$.
Then
$$\displaylines{
 \Phi_1 = U \Phi_2 =  \sum_{n_1, n_2, n_3\ge 1} c_{n_1, n_2, n_3}  
U |u^1_{n_1} u^2_{n_2} u^3_{n_3}\>
=\sum_{m=1}^M b_m U|\varphi^1_m \varphi^2_m \varphi^3_m\>
\crh =\sum_{m=1}^M b_m |\hat \varphi^1_m \hat \varphi^2_m \hat \varphi^3_m\>  
}$$
and
$$\<\psi^i_k|\hat \varphi^i_m\>_U = \<\psi^i_k u^2_{n_2}
u^3_{n_3}| U^*|\hat \varphi^i_m \hat u^2_{n_2} \hat u^3_{n_3}\>
= \<\psi^i_k u^2_{n_2}
u^3_{n_3}|\varphi^i_m u^2_{n_2} u^3_{n_3}\>
= \<\psi^i_k|\varphi^i_m\>.$$

Thus $\Phi_1$ has the expansions in $\H'_1 \otimes \H'_2 \otimes
\H'_3$ that $\Phi_2$ has in $\H_1 \otimes \H_2 \otimes \H_3$. (3.9) and
(3.10) follow.  \hfill $\blacksquare$

As before, the argument in this proof can be applied to produce a vast
range of different finite expansions.
\medskip

\proclaim{4. Isolation.}
\endproclaim

Clifton (1994) showed that a dense set of $\Psi$ do not have a triorthogonal
decomposition of the form defined by theorem 2.3, and he gave examples
of wavefunctions with no decomposition of the form defined in theorem
2.1.  He argued that this was enough to make the theorems irrelevant as
solutions to the problems of the modal interpretation caused by
degeneracies in the Schmidt decomposition.  The non-universality of
triorthogonal decompositions was also pointed out by Peres (1995).  In this
section, I shall use continuity arguments to show that there are open sets of
wavefunctions containing no elements with various decompositions.  I shall
also use a thermodynamic argument to suggest that triorthogonal decompositions
are not relevant to the measurement problem.

For a state (density matrix) $\rho$ on a Hilbert space $\H$, let
$(r_n(\rho))_{n\geq 1}$ be the unique complete decreasing sequence of
eigenvalues of $\rho$, allowing repetitions.  Thus, $\rho$ takes the form $\rho =
\sum_{n \geq 1} r_n(\rho) |\psi_n\>\<\psi_n|$ for some orthonormal basis $(\psi_n)_{n
\geq 1}$ of $\H$ with $1 \geq r_1(\rho) \geq r_2(\rho) \geq \dots \geq 0$.  The entropy
$S(\rho)$ is defined by $S(\rho) = - \tr(\rho \log \rho) = - \sum_{n\geq1} r_n(\rho) \log
r_n(\rho)$. 

\proclaim{Lemma}{\sl}
\item{\bf 4.1} For each fixed $n$, $r_n(\rho)$ is continuous in $\rho$.

\item{\bf 4.2}  $0 \leq S(\rho) \leq \log (\dim \H)$.

\item{\bf 4.3} On a finite-dimensional space, $S$ is continuous.

\item{\bf 4.4} On an infinite-dimensional space, $S$ is lower semicontinuous. 
Thus, at the limit of a convergent sequence, $S$ can jump down but not up.

\endproclaim

\proof  This is all quite standard.  A proof of (4.1) can be found in lemma
(2.1) of Bacciagaluppi, Donald,  and Vermaas (1995),
following a remark of Simon (1973).  The lower bound in (4.2) holds
because each term in the sum is positive.  The upper bound can be proved
using the non-positivity (or,  according to convention, non-negativity) of
relative entropy. This important inequality says that $\tr(-\rho \log \rho + \rho
\log \omega) \leq 0$ for all states $\omega$.  Choosing for $\omega$ the completely mixed
state yields the upper bound of (4.2).  (4.3) and (4.4) follow from (4.1).  On
a finite-dimensional space, $S(\rho)$ is a finite sum of continuous functions. 
On an infinite-dimensional space, $S(\rho)$ is the supremum of the family of
continuous non-negative functions corresponding to the partial sums.
\hbox{\ }\hfill $\blacksquare$
\medskip

If $\H = \H_1 \otimes \H_2 \otimes \H_3$, then let $\rho_i$ (respectively
$\rho_{ij}$) denote the partial trace of $\rho$ on $\H_i$ (resp. on $\H_i \otimes
\H_j$).

\proclaim{Lemma 4.5}{\sl}  Suppose that $\Psi \in \H_1 \otimes \H_2
\otimes \H_3$ has a decomposition of the form
$\Psi = \sum_{k = 1}^K a_k \psi^1_k \psi^2_k \psi^3_k$ where $K$ is finite.

Then $S((|\Psi\>\<\Psi|)_i) \leq \log K$ for $i = 1, 2, 3$.
\endproclaim

\proof  Using the Gram-Schmidt orthogonalization process, it is possible to
find an orthonormal basis $(u^i_n)_{n\geq1}$ for $\H_i$ such that $\{\psi^i_k :
k = 1, \dots, K\}$ is in the span of $\{u^i_k : k = 1, \dots, K\}$.

Then $(|\Psi\>\<\Psi|)_i$ is a density matrix on the $K$-dimensional Hilbert
space spanned by $\{u^i_k : k = 1, \dots, K\}$ and so the result is a
consequence of (4.2).
\hfill $\blacksquare$
\medskip

\proclaim{Theorem}{\sl}

\item{\bf 4.6} Suppose that $\H = \H_1 \otimes \H_2 \otimes \H_3$
and that $\dim \H_3 > \dim \H_1$.  Then there exists $\Phi \in \H$ and $\delta
> 0$ such that $||\Phi - \Psi|| > \delta$ for any $\Psi$ with a decomposition
of the form $\Psi = \sum_{k = 1}^K a_k \psi^1_k \psi^2_k \psi^3_k$ where
$\{\psi^1_k: k = 1, \dots, K\} \subset \H_1$ is a linearly independent set of
wavefunctions.
\smallskip

\item{\bf 4.7}  Suppose that $\H = \H_1 \otimes \H_2 \otimes \H_3 \otimes \H_4$
with  $\dim \H_m = N \geq 2$ for $m = 1, 2, 3, 4$.  Then there exists $\Phi \in
\H$ and $\delta > 0$ such that $||\Phi - \Psi|| > \delta$ for any $\Psi$ with a
decomposition of the form
$\Psi = \sum_{k = 1}^K a_k \psi^1_k \psi^2_k \psi^3_k \psi^4_k$ where at
least one of the sets $\{\psi^m_k: k = 1, \dots, K\}$ is linearly independent.

\endproclaim

\proof 

For $i = 1, 2, 3$, let $(u^i_n)_{n=1}^{N_i}$ be an
orthonormal basis for $\H_i$. 

\noindent{(4.6)} \  In a Hilbert space of dimension $N$, any set of linear
independent vectors can have at most $N$ elements. Thus, by lemma 4.5, if
$\Psi$ has a decomposition of the required form, 
 then $S((|\Psi\>\<\Psi|)_3) \leq \log N_1$.

 Let 
$$\Phi = \left(\sum_{n = 1}^{N_1} {1 \over \sqrt{N_1 + 1}}u^1_n u^2_1 u^3_n
\right) +  {1 \over \sqrt{N_1 + 1}} u^1_1 u^2_2 u^3_{N_1 + 1}.$$

Then $(|\Phi\>\<\Phi|)_3 = \dsize{1 \over {N_1 + 1}}\sum_{n =
1}^{N_1+1}|u^3_n\>\<u^3_n|$ so that
$S((|\Phi\>\<\Phi|)_3) = \log (N_1 + 1)$.

As $S$ is lower semicontinuous, there exists $\delta > 0$ such that
$||\Phi - \Phi'|| < \delta \Rightarrow S((|\Phi'\>\<\Phi'|)_3) \geq \log (N_1
+  {1\over 2})$.
\medskip

\noindent{(4.7)} \  Let $(u^4_n)_{n=1}^{N}$ be an orthonormal basis for
$\H_4$.  Applying lemma 4.5 to the triple product $\H = \H_1 \otimes (\H_2
\otimes \H_3) \otimes \H_4$ shows that, if $\Psi$ has a decomposition of the
required form, then $S((|\Psi\>\<\Psi|)_{23}) \leq \log K$.  Linear independence
implies that $K \leq N$.

Let 
$$\Phi = \left(\sum_{n = 1}^{N} {1 \over \sqrt{N + 1}}u^1_n u^2_n u^3_1
u^4_1 \right) +  {1 \over \sqrt{N + 1}} u^1_1 u^2_1 u^3_2 u^4_2.$$

Then $(|\Phi\>\<\Phi|)_{23} = \dsize{1 \over {N + 1}}\left(\sum_{n =
1}^{N}|u^2_n u^3_1\>\<u^2_n u^3_1| +
|u^2_1 u^3_2\>\<u^2_1 u^3_2|\right) $ so that
\newline $S((|\Phi\>\<\Phi|)_{23}) = \log (N + 1)$.
\hfill $\blacksquare$
\medskip

This theorem confirms the intuition that a small system provides too few
degrees of freedom to allow a spanning set of wavefunctions to be chosen
in each situation to correlate with product wavefunctions on a
sufficiently large product system.  The argument allows many variations.  For
example, the argument of (4.6) also proves that if $\dim \H_3 > \dim \H_1$ and
$\dim \H_3 > \dim \H_2$, then there exists $\Phi \in \H$ and $\delta > 0$ such
that $||\Phi - \Psi|| > \delta$ for any $\Psi$ with a decomposition of the form
$\Psi = \sum_{k = 1}^K a_k \psi^1_k \psi^2_k \psi^3_k$ where either
$\{\psi^1_k: k = 1, \dots, K\} \subset \H_1$ or $\{\psi^2_k: k = 1, \dots, K\}
\subset \H_2$ is a linearly independent set of wavefunctions.  This gets round
the variant of theorem 2.1 in which $\{\psi^1_k: k = 1, \dots, K\}$ need not
be linearly independent.  (4.7) extends immediately to products of more than
four spaces.

Nevertheless, in spaces with dimensions not satisfying the conditions of the
theorem or its variants, dense sets of wavefunctions satisfying the conditions
of theorem 2.2 may exist.  This holds, for example, for triple products of
infinite-dimensional spaces by theorem 3.4.  It also holds for triple products
of two-dimensional spaces.  Ac{\'\i}n et al.~(2000) show that a dense set in
such a space can be represented as a superposition of two triple-product
wavefunctions, and, as we have seen above, small variations will make the
elements of such products linearly independent.

\proclaim{Lemma 4.8}{\sl}  Let $\Psi \in \H_1 \otimes \H_2 \otimes \H_3$
have a triorthogonal decomposition $\Psi = \sum_{k = 1}^K a_k \psi^1_k \psi^2_k
\psi^3_k$, $|a_k| > 0$ for $k = 1, \dots, K$, and, for $i = 1,
2, 3$, $\{\psi^i_k: k = 1, \dots, K\}$ is a orthonormal set of wavefunctions in
$\H_i$.  Suppose that the terms in the decomposition are ordered so that
$|a_1| \geq |a_2| \geq \dots \geq |a_K|$.

Then  
$$r_k((|\Psi\>\<\Psi|)_1) = r_k((|\Psi\>\<\Psi|)_2) = r_k((|\Psi\>\<\Psi|)_3) =
|a_k|^2 \eqno{(4.9)}$$ for $k = 1, \dots, K$ and
$$S((|\Psi\>\<\Psi|)_1) = S((|\Psi\>\<\Psi|)_2) = S((|\Psi\>\<\Psi|)_3). 
\eqno{(4.10)}$$
\endproclaim

\proof This is immediate from the definitions.  It is not necessary
to assume that $K$ is finite.
\hfill
$\blacksquare$
\medskip

(4.9) and (4.10) place strong constraints on triorthogonal wavefunctions. 
For example, (4.9) and (4.1) show that there is a neighbourhood of the
wavefunction $\Psi$ of example 3.1 which contains no triorthogonal
wavefunctions.  More generally, let $T$ be the set of triorthogonal
wavefunctions in an arbitrary Hilbert space.  Clifton (1994) pointed out
that the complement of $T$ is a dense set, because a small perturbation
to the component wavefunctions of a triorthogonal decomposition can
change it to a decomposition which is not triorthogonal but which does
satisfy theorem 2.2.  This perturbed wavefunction is not triorthogonal
by the uniqueness of its decomposition.

Let $O$ be the complement of the closure of $T$.  By definition, $O$ is open.  
We can prove that arbitrary wavefunctions are unlikely to be triorthogonal by
proving that $O$ is also dense.  There are two ways of doing this.  In theorem
5.7, I shall prove that $T$ is closed.  It is easier, however, to show that
$O$ is dense directly, by using lemma 4.11.  This shows that there is a point
in $O$ arbitrarily close to every point in $T$.  It follows that arbitrarily
close to any wavefunction $\Phi$ there is a point in $O$, either because
$\Phi$ is itself in $O$, or because $\Phi$ is in $T$, or because $\Phi$ is in
$\overline T$ in which case $\Phi$ is arbitrarily close to wavefunctions in $T$.

\proclaim{Lemma 4.11}{\sl}  Let $\Psi$ be a triorthogonal wavefunction and
choose $\varepsilon \in (0, 1)$.  Then there exists $\Psi(\varepsilon)$ with 
$||\Psi - \Psi(\varepsilon)||^2 \leq 2\varepsilon$ such that no wavefunction
in a neighbourhood of $\Psi(\varepsilon)$ is triorthogonal.
\endproclaim

\proof  There are two cases to be considered.  Write $\eta = \sqrt{1 - \varepsilon}$
and $\eta' = \sqrt{\varepsilon}$.

\noindent I)  Suppose $\Psi = \psi^1_1 \psi^2_1 \psi^3_1$.  Then let
$$\Psi(\varepsilon) = \eta \psi^1_1 \psi^2_1 (\eta \psi^3_1 + \eta'\psi^3_2) +
\eta' \psi^1_2  \psi^2_2 \psi^3_2$$
where $\psi^i_1$ and $\psi^i_2$ are orthogonal wavefunctions for $i = 1, 2, 3$.
\medskip

\noindent II) Suppose $\Psi$ has a
triorthogonal decomposition of the form
$\Psi = \sum_{k = 1}^K a_k \psi^1_k \psi^2_k \psi^3_k$ where 
$1 > |a_1| \geq |a_2| \geq \dots \geq |a_K|$.  In this case $|a_2| > 0$.

Set
$$\Psi(\varepsilon) =  a_1\psi^1_1 \psi^2_1 (\eta \psi^3_1 + \eta'\psi^3_2) +
a_2 \psi^1_2  \psi^2_2 \psi^3_2 + \textstyle
\sum_{k= 3}^K a_k \psi^1_k  \psi^2_k \psi^3_k$$ where
$\eta = \sqrt{1 - \varepsilon}$ and $\eta' = \sqrt{\varepsilon}$.

In both cases, it is straightforward to calculate $||\Psi - \Psi(\varepsilon)||^2$ and
\newline $r_1(|\Psi(\varepsilon)\>\<\Psi(\varepsilon)|_i)$ for $i = 1, 2, 3$, and to show that 
 $||\Psi - \Psi(\varepsilon)||^2 \leq 2\varepsilon$ and that
$$r_1((|\Psi(\varepsilon)\>\<\Psi(\varepsilon)|)_1) = r_1((|\Psi(\varepsilon)\>\<\Psi(\varepsilon)|)_2) \ne
r_1((|\Psi(\varepsilon)\>\<\Psi(\varepsilon)|)_3).$$

The result then follows from 4.1 and 4.9. \hfill $\blacksquare$
\medskip

Using (4.10) and equating von Neumann entropy with thermodynamic
entropy would seem to confirm that triorthogonal decompositions do not exist
for typical systems consisting of measured objects, measuring devices, and
environments.  This may, however, appear a slightly curious argument to
make in a paper which criticizes as unphysical another mathematical
structure on the grounds that it is unstable, because entropy itself is also
unstable in infinite dimensional spaces.  More precisely, if $\rho$ is any state
on an infinite dimensional Hilbert space $\H$, then, in any neighbourhood of
$\rho$ there exists a state $\rho'$ with $S(\rho') = \infty$.  $\rho'$ can be constructed
by mixing with $\rho$ an arbitrarily small amount of any given state with
infinite entropy and using the concavity of $S$.  Nevertheless, from a physical
point of view, the instability of $S$ is irrelevant because what is important
physically is the minimization of local free energy.  Along similar lines, it
is conceivable that a physically-motivated algorithm could be invoked which
would imply that only states with appropriate tridecompositions are physically
relevant.  I think it implausible that such an idea could work, but the
proposals of Spekkens and Sipe (2000) might suggest a starting point.

A triorthogonal decomposition will in general fail to exist because a
macroscopic environment will occupy far more degrees of
freedom than a microscopic object.  Free energy minimization constrains the
local quantum state to an effectively finite dimensional space of bounded local
energy in which entropy is a physically-revelant finite measure of the number
of available degrees of freedom.  The approach to local equilibrium is a
process of exploring the available state space in a way that rules out the
sort of wavefunction correlation expressed by a triorthogonal
decomposition unless equation (4.10) happens to hold.  But local equilibrium
equalizes local temperatures rather than local entropies.

In a paper which is ultimately about collapse or the appearance of collapse,
it may seem that an appeal to statistical equilibrium is also rather curious. 
Perfect statistical equilibrium allows only thermodynamic parameters to be
observed.  Nevertheless, it seems to me that the states of quantum
statistical mechanics almost always provide a better approximation to the
correct description of observed macroscopic objects than do the
wavefunctions of elementary quantum mechanics, given how little
information we usually have about such objects.

Equating von Neumann entropy with thermodynamic entropy does
more for us than merely to rule out triorthogonal decompositions.  For
bipartite systems, a Schmidt decomposition implies equality of von
Neumann entropy for the two subsystems.  Since a Schmidt decomposition
exists for all bipartite pure states, in my view, this implies that the answer
to question 1.2 is that it is almost always inappropriate to assume that the
state of any macroscopic bipartite system, except possibly the entire universe,
is pure.   The same conclusion follows if we just equate von Neumann entropy
with thermodynamic entropy for the total system.  If observed systems have
non-zero entropy then they should not be described by pure states.

\proclaim{5. Stability for Triorthogonal Decompositions.}
\endproclaim

We turn in this section to a detailed analysis of decompositions
satisfying theorem 2.3.  In theorem 5.6, it will be shown that triorthogonal
decompositions are not only unique, but also stable, and in theorem 5.7,
it will be shown that the set $T$ of triorthogonal wavefunctions is closed. 
Given the consequent rarity of triorthogonal wavefunctions, I suspect that the
main interest in this section lies in the methods used and in lemma 5.5,
rather than in these theorems.

The proofs in this section depend on careful estimations using the
following three lemmas.  These are essentially standard.  Recall that $||\
||_1$ denotes the trace norm.   Reed and Simon (1972, chapter VI)
provides an introduction to trace class operators.

\newpage

\proclaim{Lemma 5.1}{\sl}  Let $R = \sum_n r_n |\psi_n\>\<\psi_n|$ and $S  =
\sum_n s_n |\varphi_n\>\<\varphi_n|$ be eigenvalue decompositions of positive
trace class operators with
$1 \geq r_1 \geq \dots \geq r_n \geq \dots \geq 0$ and
 $1 \geq s_1 \geq \dots \geq s_n \geq \dots \geq 0$.
Then $|r_n - s_n| \leq ||R - S||_1$ for $n = 1, 2, \dots$.
\endproclaim

\proof  This is lemma 2.1 of Bacciagaluppi, Donald,  and Vermaas (1995)
without the restriction that $\tr(R) = \tr(S) = 1$ which is not needed for the
proof. \hfill $\blacksquare$

\proclaim{Lemma}{\sl} Let $\Psi$ and $\Phi$ be wavefunctions and $P$ be
a projection.  Then  
$$||P(|\Psi\>\<\Psi| - |\Phi\>\<\Phi|)P||_1 \leq ||\,|\Psi\>\<\Psi| - |\Phi\>\<\Phi|\,||_1 \leq
2||\Psi - \Phi|| \eqno{(5.2)}$$ 
If, moreover, $\<\Psi|\Phi\> > 0$, then
$$||\,|\Psi\>\<\Psi| - |\Phi\>\<\Phi|\,||_1 \leq 2||\Psi - \Phi|| \leq
\sqrt{2}||\,|\Psi\>\<\Psi| - |\Phi\>\<\Phi|\,||_1 \eqno{(5.3)}$$
\endproclaim

\proof  The first inequality of (5.2) follows from the general result that if
$A$ is a trace class operator and $B$ is bounded, then
$||AB||_1 \leq ||A||_1 ||B||$ and $||BA||_1 \leq ||A||_1 ||B||$.

For wavefunctions $\Psi$ and $\Phi$,  $||\Psi - \Phi||^2 = 2 - \<\Psi|\Phi\> -
\<\Phi|\Psi\> \geq 2(1 - |\<\Psi|\Phi\>|)$. 
Explicit diagonalization of
$|\Psi\>\<\Psi| - |\Phi\>\<\Phi|$ gives
$$||\,|\Psi\>\<\Psi| - |\Phi\>\<\Phi|\,||_1 = 2\sqrt{1 - |\<\Psi|\Phi\>|^2}.$$

Thus
$$\displaylines{
 ||\,|\Psi\>\<\Psi| - |\Phi\>\<\Phi|\,||_1 = 2\sqrt{1 - |\<\Psi|\Phi\>|} \sqrt{1 +
|\<\Psi|\Phi\>|}
\cr \leq 2\sqrt{2(1 - |\<\Psi|\Phi\>|)}  \leq 2||\Psi - \Phi||.
}$$

If $\<\Psi|\Phi\> > 0$, then
$$||\Psi - \Phi||^2 = 2 - \<\Psi|\Phi\> - \<\Phi|\Psi\> = 2(1 - |\<\Psi|\Phi\>|)$$ and so
$$\displaylines{
 ||\,|\Psi\>\<\Psi| - |\Phi\>\<\Phi|\,||_1 = 2\sqrt{1 - |\<\Psi|\Phi\>|} \sqrt{1 +
|\<\Psi|\Phi\>|}
 \geq \sqrt{2}||\Psi - \Phi||.
}$$
\hfill $\blacksquare$

\proclaim{Lemma 5.4}{\sl}  If $A$ is a trace class operator on a tensor
product Hilbert space $\H = \H_1 \otimes \H_2$, and $A_1$ is the partial
trace of $A$ on $\H_1$, then $||A_1||_1 \leq ||A||_1$.
\endproclaim

\proof  A proof of this is provided in lemma 4.7 of Bacciagaluppi, Donald, 
and Vermaas (1995). \hbox{\ } \hfill $\blacksquare$
\medskip

At the heart of theorem 2.3 lies the fact that a product wavefunction
$\psi^1\psi^2$ cannot be decomposed into a non-trivial sum of
products.  The next lemma develops that fact by providing explicit bounds. 
In this lemma, it is not assumed that $\Psi$ or $\Phi$ are normalized.

\proclaim{Lemma 5.5}{\sl}  Let $\Psi = a \psi^1 \psi^2$ and
$\Phi = \sum_k b_k \varphi^1_k \varphi^2_k$ where $\psi^1$ and $\psi^2$ are
wavefunctions and, for $i = 1, 2$, $(\varphi^i_k)_k$ are orthonormal sequences
of wavefunctions. 

Suppose that $||(|\Psi\>\<\Psi|)_1 - (|\Phi\>\<\Phi|)_1||_1 < \varepsilon'$ and that
 $1 \geq |a|^2 > 2\varepsilon' > 0$.

Then there exists a unique $m$ such that
$|\, |a|^2 - |b_m|^2|< \varepsilon'$ and $|b_{m'}|^2 < \varepsilon'$ for $m \ne m'$.
\medskip

Suppose further that $||\Psi - \Phi|| < \varepsilon'$ and that
$\sqrt{\varepsilon'} \leq |a|\varepsilon/3 < 1$ for some $\varepsilon \in (0,
1)$.

Then $||a\psi^1\psi^2 - b_m\varphi^1_m\varphi^2_m|| < \varepsilon$,
$|\<\psi^1|\varphi^1_m\>| > 1 - \varepsilon$, and $|\<\psi^2|\varphi^2_m\>| > 1 -
\varepsilon$.
\endproclaim

\proof  Suppose that $||(|\Psi\>\<\Psi|)_1 - (|\Phi\>\<\Phi|)_1||_1 < \varepsilon'$.

 $(|\Psi\>\<\Psi|)_1 = |a|^2 |\psi^1\>\<\psi^1|$, 
$(|\Phi\>\<\Phi|)_1 = \sum_k |b_k|^2 |\varphi^1_k\>\<\varphi^1_k|$.

Choose $m$ such that $|b_m| = \max\{ |b_k| \}$.  By lemma 5.1,
$|\, |a|^2 - |b_m|^2|< \varepsilon'$ and $|b_{m'}|^2 < \varepsilon'$ for $m' \ne m$.

If $|a|^2 > 2\varepsilon'$ and $|b_{m'}|^2 < \varepsilon'$, then
$|\, |a|^2 - |b_{m'}|^2| = |a|^2 - |b_{m'}|^2 > \varepsilon'$ and so $m$ is unique.

Now suppose also that $||\Psi - \Phi|| < \varepsilon'$ and that
$\sqrt{\varepsilon'} \leq |a|\varepsilon/3 < 1$, for some $\varepsilon \in (0,
1)$. 

Write $\Phi' = b_m\varphi^1_m\varphi^2_m$ and $\Phi'' = \Phi - \Phi' = \sum_{k \ne m}
b_k\varphi^1_k\varphi^2_k$.

By orthogonality, $||\Phi||^2 = ||\Phi'||^2 + ||\Phi''||^2$.

$||\Psi|| = |a|$ and $||\Phi'|| = |b_m|$ so we know that $|\,||\Psi||^2 -
||\Phi'||^2\,| < \varepsilon'$.  Thus
$$\displaylines{
 ||\Phi''||^2 = ||\Phi||^2 - ||\Phi'||^2 = ||\Phi||^2 - ||\Psi||^2 + ||\Psi||^2 - ||\Phi'||^2
\cr = (||\Phi|| - ||\Psi||)(||\Phi|| + ||\Psi||) + |a|^2 - |b_m|^2
\cr \leq (||\Psi - \Phi||)(2||\Psi|| + ||\Phi - \Psi||) + \varepsilon' \leq \varepsilon'(2|a| + \varepsilon') + \varepsilon' \leq
4\varepsilon'   }$$
and
$$\displaylines{
 ||a\psi^1\psi^2 - b_m\varphi^1_m\varphi^2_m|| = ||\Psi - \Phi + \Phi''|| \leq ||\Psi - \Phi|| +
||\Phi''||  <  \varepsilon' + 2\sqrt{\varepsilon'} < 3\sqrt{\varepsilon'}
\leq \varepsilon.
}$$

$$||\Psi||\, ||\Psi - \Phi|| \geq |\<\Psi|\Psi - \Phi\>| = |\,||\Psi||^2 - \<\Psi|\Phi\>|
\geq ||\Psi||^2 - |\<\Psi|\Phi\>|$$
so that $|\<\Psi|\Phi\>| \geq ||\Psi||^2 - ||\Psi||\, ||\Psi - \Phi|| >
|a|(|a| - \varepsilon')$.

But $\<\Psi|\Phi\> = \bar a (b_m \<\psi^1|\varphi^1_m\> \<\psi^2| \varphi^2_m\> +
\sum_{k\ne m} b_k \<\psi^1|\varphi^1_k\> \<\psi^2| \varphi^2_k\>)$ and so 
$$\displaylines{
 |\<\Psi|\Phi\>| \leq |a| (|b_m| |\<\psi^1|\varphi^1_m\>| |\<\psi^2|\varphi^2_m\>|
 +
\varepsilon'
\textstyle\sqrt{\sum_k |\<\psi^1|\varphi^1_k\>|^2} \sqrt{\sum_k |\<\psi^2|\varphi^2_k\>|^2}\ ) 
\crh \leq |a| (|b_m| |\<\psi^1|\varphi^1_m\>| |\<\psi^2|\varphi^2_m\>| + \varepsilon') 
}$$

Thus, for $|a| > 0$,
$|b_m| |\<\psi^1|\varphi^1_m\>| |\<\psi^2|\varphi^2_m\>| > |a| -
2\varepsilon'$.

$$\displaylines{
 |a| |\<\psi^1|\varphi^1_m\>| |\<\psi^2|\varphi^2_m\>|
\hcr = |b_m| |\<\psi^1|\varphi^1_m\>| |\<\psi^2|\varphi^2_m\>| + (|a| - |b_m|)
|\<\psi^1|\varphi^1_m\>| |\<\psi^2|\varphi^2_m\>| \crh > |a| - 2\varepsilon' 
- |\,|a| - |b_m|\,|.   }$$

Now $\varepsilon' > |\,|a|^2 - |b_m|^2\,|$ implies that 
$\varepsilon' > |\,|a| - |b_m|\,|(|a| + |b_m|) \geq |\,|a| - |b_m|\,|^2$ and
so $\sqrt{\varepsilon'} > |\,|a| - |b_m|\,|$ and
$$|a| |\<\psi^1|\varphi^1_m\>| |\<\psi^2|\varphi^2_m\>|
> |a| - 2\varepsilon'  - \sqrt{\varepsilon'} > |a| - 3\sqrt{\varepsilon'}.$$

Using $\sqrt{\varepsilon'} \leq |a|\varepsilon/3 < 1$  yields
$|a| |\<\psi^1|\varphi^1_m\>| |\<\psi^2|\varphi^2_m\>| > |a|(1 - \varepsilon)$
and so $|\<\psi^i|\varphi^i_m\>| > 1 - \varepsilon$ for $i = 1, 2$.  \hfill
$\blacksquare$
\medskip

The aim now is to analyse the relationship between the components of two
neighbouring wavefunctions $\Psi$ and $\Phi$ with triorthogonal
decompositions $$\Psi = \sum_{k = 1}^K a_k \psi^1_k \psi^2_k \psi^3_k
\quad \hbox{ and } \quad \Phi = \sum_{k = 1}^{K'} b_k \varphi^1_k \varphi^2_k
\varphi^3_k.$$ 

We shall call a triorthogonal decomposition $\Psi = \sum_{k = 1}^K a_k \psi^1_k
\psi^2_k \psi^3_k$ an ``ordered triorthogonal decomposition'' if the 
$|a_k|$ are non-increasing ($|a_1| \geq |a_2| \geq \dots \geq 0$).  We may also
write the decomposition in the form
 $\Psi = \sum_{m = 1}^M \hat a_m (\sum_{k = 1}^{K_m} \psi^1_{mk}
\psi^2_{mk}\psi^3_{mk})$, where $(|\hat a_m|)_{m=1}^M$ is the strictly
decreasing sequence of distinct non-zero values for the $|a_k|$ ($|\hat a_1|
> |\hat a_2| > \dots > 0$).  We shall call this a ``strictly ordered triorthogonal
decomposition''.

  The complexities of theorem 5.6 and its proof arise
because the phases of the component wavefunctions are not determined by
the decomposition; because the sums may not be finite; and because the
smaller in absolute value the coefficients
$a_k$ and $b_k$, the less the corresponding components need agree for a
given difference between the total wavefunctions.

\proclaim{Theorem 5.6}{\sl}  Let 
$\dsize \Psi = \sum_{m = 1}^M \hat a_m (\sum_{k = 1}^{K_m} \psi^1_{mk}
\psi^2_{mk} \psi^3_{mk})$ be a wavefunction with a strictly ordered
triorthogonal decomposition.  Choose a finite integer $L$ with $1 \leq L \leq M$
and $\varepsilon \in (0, {1\over 4})$.  Suppose that $\Phi$ is a wavefunction with a
triorthogonal decomposition of the form  $\Phi = \sum_{k = 1}^{K'} b_k \varphi^1_k
\varphi^2_k
\varphi^3_k$ and that  $||\Psi - \Phi|| < |\hat a_L|^2 \varepsilon^2/18$.  

Then, for each $m \in \{1, 2, \dots L\}$ and $k \in \{1, \dots, K_m\}$, there
exists a unique $k' \in \{1, 2, \dots K'\}$  such that
$ |\, |\hat a_m|^2 - |b_{k'}|^2| < 3\varepsilon$, such that
$|\<\psi^i_{mk}|\varphi^i_{k'}\>| > 1 - \varepsilon$  for $i = 1, 2, 3$, and such that
$||\hat a_m \psi^1_{mk} \psi^2_{mk}\psi^3_{mk} -
b_{k'} \varphi^1_{k'}\varphi^2_{k'}\varphi^3_{k'}||  < 3\sqrt{\varepsilon}$.
\endproclaim

\proof  
Set $\varepsilon' = |\hat a_L|^2 \varepsilon^2/9$.  Then, for $m = 1, 2, \dots, L$, $|\hat
a_m|^2 > 2\varepsilon' > 0$ and $\varepsilon' < \sqrt{\varepsilon'} \leq |\hat
a_m|\varepsilon/3 < 1$.

Let $P_{\psi^3_{mk}}$ be the orthogonal projection from $\H_1 \otimes \H_2
\otimes \H_3$ onto $\H_1 \otimes \H_2 \otimes \{\psi^3_{mk}\}$.

Then $P_{\psi^3_{mk}}\Psi = \hat a_m \psi^1_{mk} \psi^2_{mk} \psi^3_{mk}$ and
$$P_{\psi^3_{mk}}\Phi = \sum_{k' = 1}^{K'} b_{k'}\<\psi^3_{mk}|\varphi^3_{k'}\>
\varphi^1_{k'}\varphi^2_{k'} \psi^3_{mk}.$$

Write $R = |P_{\psi^3_{mk}}\Psi\>\<P_{\psi^3_{mk}}\Psi|$ and 
$S = |P_{\psi^3_{mk}}\Phi\>\<P_{\psi^3_{mk}}\Phi|$.

Then, by (5.2) and lemma 5.4, $||\Psi - \Phi|| <  {1\over 2}\varepsilon'
\Rightarrow ||R_1 - S_1||_1 <
\varepsilon'$.  Also, of course, $||P_{\psi^3_{mk}}\Psi - P_{\psi^3_{mk}}\Phi|| <  {1\over 2}\varepsilon' < \varepsilon'$
and so, by lemma 5.5, there is a unique $k'$ such that 
$$|\, |\hat a_m|^2 - |b_{k'}|^2|\<\psi^3_{mk}|\varphi^3_{k'}\>|^2|< \varepsilon',$$ such that
$$||\hat a_m \psi^1_{mk} \psi^2_{mk} -
 b_{k'}\<\psi^3_{mk}|\varphi^3_{k'}\> \varphi^1_{k'}\varphi^2_{k'} || < \varepsilon,$$ and 
such that
$|\<\psi^1_{mk}|\varphi^1_{k'}\>| > 1 - \varepsilon$ and $|\<\psi^2_{mk}|\varphi^2_{k'}\>| > 1 -
\varepsilon$.

As $1 - \varepsilon > {1\over \sqrt{2}}$, $\varphi^1_{k'}$ and $\varphi^2_{k'}$ are
uniquely determined by these inequalities.  Projecting onto
$\{\psi^1_{mk}\} \otimes \H_2 \otimes \H_3$ implies that
$|\<\psi^3_{mk}|\varphi^3_{k'}\>| > 1 - \varepsilon$ and so
$$\displaylines{
 |\, |\hat a_m|^2 - |b_{k'}|^2|
 \leq |\, |\hat a_m|^2 - |b_{k'}|^2|\<\psi^3_{mk}|\varphi^3_{k'}\>|^2|
+ |b_{k'}|^2(1-|\<\psi^3_{mk}|\varphi^3_{k'}\>|^2)
\hcrh < \varepsilon' + 2(1-|\<\psi^3_{mk}|\varphi^3_{k'}\>|) \leq 3\varepsilon.  
}$$

For all wavefunctions $\varphi$ and $\psi$, $||\varphi - \<\psi|\varphi\> \psi||^2 = 1 - 
|\<\psi|\varphi\>|^2$.  It follows that
$$\displaylines{
 ||b_{k'} \varphi^1_{k'}\varphi^2_{k'}\varphi^3_{k'} - b_{k'}\<\psi^3_{mk}|\varphi^3_{k'}\>
\varphi^1_{k'}\varphi^2_{k'} \psi^3_{mk}||^2 = |b_{k'}|^2(1 -
|\<\psi^3_{mk}|\varphi^3_{k'}\>|^2) \leq 2\varepsilon 
}$$
and so
$$\displaylines{
 ||\hat a_m \psi^1_{mk} \psi^2_{mk}\psi^3_{mk} -
b_{k'} \varphi^1_{k'}\varphi^2_{k'}\varphi^3_{k'}|| 
\hcrh \leq  ||\hat a_m \psi^1_{mk} \psi^2_{mk}\psi^3_{mk} -
 b_{k'}\<\psi^3_{mk}|\varphi^3_{k'}\> \varphi^1_{k'}\varphi^2_{k'}
\psi^3_{mk}|| + 2\sqrt{\varepsilon} < 3\sqrt{\varepsilon}. }$$
\hfill $\blacksquare$

\proclaim{Theorem 5.7}  Let $(\Psi_n)_{n\geq1}$ be a sequence of triorthogonal
wavefunctions and suppose that $\Psi_n \rightarrow \Psi$.  Then $\Psi$ is triorthogonal.
\endproclaim

\proof  Choose $\varepsilon \in (0, {1 \over 4})$.

Suppose that $\Psi_n = \sum_{k=1}^{K_n} a_k(n) \psi^1_k(n) \psi^2_k(n)
\psi^3_k(n)$ is a triorthogonal decomposition. 

By 5.1, 5.2, and 5.4, for $i = 1, 2, 3$ and all $k$,
$$r_k((|\Psi_n\>\<\Psi_n|)_i) \rightarrow r_k((|\Psi\>\<\Psi|)_i)$$ as $n \rightarrow \infty$.

It follows that $$r_k((|\Psi\>\<\Psi|)_1) = r_k((|\Psi\>\<\Psi|)_2) =
r_k((|\Psi\>\<\Psi|)_3).$$

Set $r_k = r_k((|\Psi\>\<\Psi|)_1)$.  

  Let $(\hat r_m)_{m = 1}^M$ be the
ordered sequence of distinct decreasing values for the non-zero $r_k$. 
For any finite integer $L \leq M$, let $T_L$ be the total number of $r_k$
greater than or equal to $\hat r_L$.

Choose $L$ such that $\sum_{k=1}^{T_L} r_k \geq 1 - \varepsilon^2$.

If $L = M$, write $\hat r_{M +1} = 0$.
Let $\alpha = \min\{ |\hat r_k - \hat r_{k+1}|: 1 \leq k \leq L\}$.

$\alpha > 0$.  There exists $N_1$ such that $n \geq N_1$ implies $||\Psi_n - \Psi|| <
\alpha/6$.

By 5.1, 5.2, and 5.4, for $n \geq N_1$
$$\displaylines{
 ||(|\Psi\>\<\Psi|)_1 - (|\Psi_n\>\<\Psi_n|)_1||_1 < \alpha/3 \cr
\hbox{and \ \ }
 |r_k((|\Psi\>\<\Psi|)_1) - r_k((|\Psi_n\>\<\Psi_n|)_1)| < \alpha/3 \hbox{\
\ for all $k$.} }$$ 

This implies that there are exactly $T_L$ values of
$r_k((|\Psi_n\>\<\Psi_n|)_1)$ which are greater than or equal to
$\hat r_L - \alpha/3$.

Now set $\varepsilon' = |\hat r_L - \alpha/3|^2 \varepsilon^2/36\}$. 

There exists $N_2 \geq N_1$ such that $n \geq N_2$ implies $||\Psi_n - \Psi|| <
\varepsilon'$, and so $m, n \geq N_2$ implies $||\Psi_m - \Psi_n|| < 2\varepsilon'$.

By theorem 5.6, there exists a unique matching of the first $T_L$ terms in
the triorthogonal decompositions of $\Psi_m$ and $\Psi_n$ such that, using
an ordering compatible with this matching, 
$|\<\psi^i_k(m)|\psi^i_k(n)\>| > 1 - \varepsilon$  for $i = 1, 2, 3$,
$ |\, |a_k(m)|^2 - |a_k(n)|^2| < 3\varepsilon$, and 
$$ ||a_k(m) \psi^1_k(m) \psi^2_k(m)\psi^3_k(m) -
a_k(n) \psi^1_k(n) \psi^2_k(n)\psi^3_k(n)||  < 3\sqrt{\varepsilon}.$$

It follows that the sequences $(|a_k(n)|)_{n\geq1}$,
$(|\psi^i_k(n)\>\<\psi^i_k(n)|)_{n\geq1}$, and \newline $(a_k(n) \psi^1_k(n)
\psi^2_k(n)\psi^3_k(n))_{n\geq1}$ are Cauchy, and hence convergent.

With a suitable choice of labelling, the limit of the
$|a_k(n)|$ can be taken to $\sqrt{r_k}$.  $|\psi^i_k(n)\>\<\psi^i_k(n)|$
converges to a pure state. Write $|\psi^i_k(n)\>\<\psi^i_k(n)| \rightarrow
|\psi^i_k\>\<\psi^i_k|$ where some particular choice of phase is made in the
limit wavefunction.

Now choose the phases of the $\psi^i_k(n)$ so that $\<\psi^i_k(n)|\psi^i_k\>
\geq 0$ for all $i$, $k$, and $n$, adjusting the phase of $a_k(n)$ to leave
$a_k(n) \psi^1_k(n) \psi^2_k(n)\psi^3_k(n)$ unchanged.

In this case, by (5.3),
$$|\psi^i_k(n)\>\<\psi^i_k(n)| \rightarrow |\psi^i_k\>\<\psi^i_k| \Rightarrow
\psi^i_k(n) \rightarrow \psi^i_k.$$

It follows that
$a_k(n) \psi^1_k(n) \psi^2_k(n)\psi^3_k(n)$ converges to $\sqrt{r_k}
e^{i\theta_k}
\psi^1_k\psi^2_k\psi^3_k$ for some phase $\theta_k$.  Setting $a_k = \sqrt{r_k}
e^{i\theta_k}$ gives
$$a_k(n) \psi^1_k(n) \psi^2_k(n)\psi^3_k(n) \rightarrow a_k \psi^1_k\psi^2_k\psi^3_k$$
and $a_k(n) \rightarrow a_k$.

Write $\Phi_L = \sum_{k=1}^{T_L} a_k \psi^1_k\psi^2_k\psi^3_k$ and let
$\Phi = \lim_{L \rightarrow M} \Phi_L$.  $\Phi$ is a triorthogonal
wavefunction.

Write $\Phi_L(n) = \sum_{k=1}^{T_L} a_k(n) \psi^1_k(n) \psi^2_k(n)\psi^3_k(n)$.

With the value of $L$ chosen earlier, $||\Phi_L - \Phi||^2 = \sum_{k > T_L} r_k
\leq \varepsilon^2$.

Using the convergence of a finite number of fixed convergent sequences,
there exists
$N_3 \geq N_2$, such that $n \geq N_3$ implies that
$$|(||\Phi_L(n) - \Psi_n|| - ||\Phi_L - \Phi||)|
= |\textstyle\sqrt{1 - \sum\limits_{k=1}^{T_L} |a_k(n)|^2} - \sqrt{1 -
\sum\limits_{k=1}^{T_L} r_k}| < \varepsilon$$ and that
$||\Phi_L(n) - \Phi_L|| < \varepsilon$.

This implies that, for $n \geq N_3$, 
$$||\Phi - \Psi|| \leq ||\Phi -\Phi_L|| + ||\Phi_L - \Phi_L(n)|| + ||\Phi_L(n) - \Psi_n||
+ ||\Psi_n - \Psi|| < 5\varepsilon.$$

As $\varepsilon$ was freely chosen, $\Psi = \Phi$ and the result is proved. \hfill $\blacksquare$

\proclaim{6. Conclusion.}
\endproclaim

An appropriate goal for a realist interpretation of quantum theory is an
algorithm providing an explanation of the appearance of collapse.  In the
modal interpretation, the proposed algorithm was based on the biorthogonal, or
Schmidt, decomposition of a wavefunction on a biproduct space $\H = \H_1
\otimes \H_2$.  Degeneracies, instabilities, and the analysis of plausible
states for thermal systems produce obstacles to this program which, in my
opinion, are insurmountable.

Bub (1997) discusses several theorems which could be useful in the
development of algorithms for realist interpretations of quantum theory. 
Theorem 2.1 is among these.  In my opinion, the instability demonstrated
in section 3 would make worthless any algorithm based purely on the
uniqueness provided by theorem 2.1.  The results in section 4, in particular
(4.10), show that theorem 2.3 is also unlikely to be of fundamental physical
significance.

The central purpose of this paper is to argue that the uniqueness provided
by the tridecompositional theorem is not a physically relevant way of
identifying component wavefunctions in product spaces.  Even if it is
thought necessary to expand wavefunctions into sums of product
wavefunctions, the conditions of theorem 2.1 may simply be too strong. 
For example, with the definitions of example 3.3, $\Psi(\theta)$ may have a
unique expansion for $\theta$ small and positive satisfying the conditions of
theorem 2.1, but, for almost all purposes, an expansion into the orthonormal
product basis $(\psi^1_{n_1} \psi^2_{n_2} \psi^3_{n_3})_{n_1, n_2, n_3}$ is
likely to be more appropriate.  This will reveal the closeness of
$\Psi(\theta)$ to $\Psi$ and show the weight in $\Psi(\theta)$ of the
correlation $\psi^1_1 \psi^2_1 \psi^3_1$.

More generally, however, I believe that it is a mistake to assume that any
physical systems should be described by pure states unless we have good
reasons to expect that such descriptions may be valid; as for example with
well-controlled and carefully prepared microscopic systems.  In particular,
I think it is a mistake to try to explain the appearance of collapse in terms
of the assumption, at any instant, of pure states for open thermal
macroscopic systems like brains or cats or measuring devices.

\proclaim{References}{}
\endproclaim

\frenchspacing
\parindent=0pt

{\everypar={\hangindent=0.75cm \hangafter=1} 

Ac{\'\i}n, A., Andrianov, A., Costa, L., Jan\'e, E., Latorre, J.I., and
Tarrach, R. (2000)  ``Generalized Schmidt decomposition and classification of
three-quantum-bit \newline states'' {\sl quant-ph/0003050}.  Printed: {\sl
Phys. Rev. Lett. \bf 85}, 1560, (2000).

Bacciagaluppi, G., Donald, M.J., and Vermaas, P.E. (1995)  ``Continuity and
discontinuity of definite properties in the modal interpretation.'' {\sl Helv.
Phys. Acta \bf 68}, 679--704.

Bub, J. (1997) {\sl Interpreting the Quantum World.} 
(Cambridge)

Cassam-Chena{\"\i} P., Patras F. (2004) ``Higher order, generalized, Schmidt
decompositions for indistinguishable, overlapping particles.''
{\sl Phys. Lett. A \bf 326}, 297--306.

Clifton, R. (1994)  ``The triorthogonal uniqueness theorem and its
irrelevance to the modal interpretation of quantum mechanics.''  In {\sl
Symposium on the Foundations of Modern Physics 1994: 70 Years of
Matter Waves} (ed. K.V. Laurikainen, C. Montonen, and K. Sunnarborg),
pages 45--60. (Editions Frontiers) 

Donald, MJ. (1998) ``Discontinuity and continuity of definite properties in
the modal interpretation.''  In {\sl The Modal Interpretation of Quantum
Mechanics} (ed. D. Diecks and P.E. Vermaas), pages 213--222.  (Kluwer)

Donald, M.J. (1999)  ``Progress in a many-minds interpretation of
quantum theory.'' {\sl quant-ph/9904001}

Donald, M.J. (2002)  ``Neural unpredictability, the interpretation of
quantum theory, and the mind-body problem.''  
{\sl quant-ph/0208033}

{\hfill My papers are also available from \ \
{\catcode`\~=12 \catcode`\q=9
http://www.poco.phy.cam.ac.uk/q~mjd1014 } \hfill}
\smallskip

Elby A. and Bub, J. (1994)  ``Triorthogonal uniqueness theorem and its
relevance to the interpretation of quantum mechanics.''  {\sl Phys. Rev. A
\bf 49}, 4213--16.

Hepp, K. (1972)  ``Quantum theory of measurement and macroscopic
observables.''  {\sl Helv. Phys. Acta \bf 45}, 237--48.

Kirkpatrick, K.A. (2001)  ``Uniqueness of a convex sum of products of
projectors.'' {\sl quant-ph/0104093}.  Printed: {\sl J. Math. Phys. \bf 43},
684--686, (2002).

Ollivier, H., Poulin, D., and Zurek, W.H. (2003)  ``Objective properties from
subjective quantum states: Environment as a witness'' {\sl quant-ph/0307229}.

Ollivier, H., Poulin, D., and Zurek, W.H. (2004)  ``Environment as a witness:
Selective proliferation of information and emergence of objectivity'' {\sl
quant-ph/0408125}.

Peres, A. (1995)  ``Higher order Schmidt decompositions.'' {\sl
quant-ph/9504006}.  Printed: {\sl Phys. Lett. A \bf 202},
16--17, (1995).

Reed, M. and Simon, B. (1972) {\sl Methods of Modern Mathematical
Physics. I: Functional Analysis.} (Academic Press)

Schlosshauer, M. (2003)  ``Decoherence, the measurement problem, and
interpretations of quantum mechanics.''  {\sl  quant-ph/0312059}.

Simon, B. (1973)  ``Convergence theorems for entropy.''  {\sl J.
Math. Phys. \bf 14}, 1940--1941.

Spekkens, R.W. and Sipe, J.E. (2000)  ``Non-orthogonal preferred projectors
for modal interpretations of quantum mechanics.''  {\sl quant-ph/0003092}. 

Vermaas, P.E. (2000) {\sl  A Philosopher's Understanding of Quantum
Mechanics: Possibilities and Impossibilities of a Modal Interpretation.} 
(Cambridge)

}

\end